\documentclass[twocolumn,letterpaper,amsmath,amssymb,floatfix,aps,superscriptaddress]{revtex4} 

\usepackage{graphicx}
\usepackage{dcolumn}
\usepackage{bm}
\usepackage{epsfig}

\def\ln{{\operatorname{ln}}}

\def\Tr{{\operatorname{Tr}}}
\def\sh{{\operatorname{sinh}}}

\def\tg{{\operatorname{tan}}}

\def\cth{{\operatorname{coth}}}
\newcommand{\Av}[1]{{\bf #1}}

\begin{document}

\title{Weak and Strong-Coupling Electrostatic Interactions between Asymmetrically Charged Planar Surfaces}

\author{M. Kandu\v c}
\affiliation{Department of Theoretical Physics,
J. Stefan Institute, SI-1000 Ljubljana, Slovenia}

\author{M. Trulsson}
\affiliation{Department of Theoretical Chemistry, Lund University
Chemical Center, P.O.B 124 S-221 00 Lund, Sweden}

\author{A. Naji}
\affiliation{Materials Research Laboratory, \&
Department of Chemistry and Biochemistry,
University of California, Santa Barbara, CA 93106}
\affiliation{Kavli Institute for Theoretical Physics, University of California, 
Santa Barbara, CA93106}

\author{Y. Burak}
\affiliation{Center for Brain Science, Harvard University, Cambridge, MASS 02138}

\author{J. Forsman}
\affiliation{Department of Theoretical Chemistry, Lund University
Chemical Center, P.O.B 124 S-221 00 Lund, Sweden}

\author{R. Podgornik}
\affiliation{Department of Theoretical Physics,
J. Stefan Institute, SI-1000 Ljubljana, Slovenia}
\affiliation{Department of Physics, Faculty of Mathematics and Physics,
University of Ljubljana, SI-1000 Ljubljana, Slovenia}
\affiliation{Laboratory of Physical and Structural Biology, NICHD, Bld. 9, Rm. 1E116, National Institutes of Health, Bethesda, MD 20892-0924}
\affiliation{Kavli Institute for Theoretical Physics, University of California, 
Santa Barbara, CA93106}

\begin{abstract}
We compare weak and strong coupling theory of counterion-mediated electrostatic interactions between two asymmetrically charged plates with extensive Monte-Carlo simulations. 
Analytical results in both weak and strong coupling limits compare excellently with simulations in their respective regimes of validity. The system shows a surprisingly 
rich structure in terms of interactions between the surfaces as well as fundamental qualitative differences in behavior in the weak and the strong coupling limits.
\end{abstract}

\maketitle

\section{Introduction}

Stability and interactions in biological and soft-matter systems often depends on the underlying properties of electrostatic interactions \cite{holm}. Charges on 
macromolecular surfaces in aqueous environments, as in the case of membranes, self-assembled micelles, globular proteins and fibrous polysaccharides, affect a 
wealth of functional, structural and dynamical properties \cite{Andelman}.
The traditional approach to charged (bio)colloidal systems has been the mean-field Poisson-Boltzmann (PB) formalism applicable at weak surface charges, low
 counter-ion valency and high temperature \cite{DLVO}. The limitations of this approach become practically important in highly-charged systems where  counterion-mediated
 interactions between charged bodies start to deviate substantially from the mean-field accepted wisdom \cite{hoda,Naji}. One of the fundamental recent advances in this
 field has been the systematization of these non-PB effects based on the notions of {\em weak} and {\em strong} coupling approximations. The latter approach has been pioneered by 
Rouzina and Bloomfield \cite{Rouzina}, elaborated later by Shklovskii {\em et al.} \cite{shklovskii}, Levin {\em et al.} \cite{Levin},  and brought into final form by 
Netz {\em et al.} \cite{Netz,hoda,Naji}. These two approximations allow for an explicit and exact treatment of charged systems at two disjoint {\em limiting conditions} 
whereas the parameter space in between can be analyzed only approximately \cite{santangelo,lue,yoram,Netz,weeks} and is mostly accessible solely {\em via} computer simulations \cite{hoda,Naji,Netz,weeks,guld84,bratko,valleau,kjelland,Jho2,trulsson}.

In the absence of a general approach that would cover thoroughly all the regions of the parameter space one has to take recourse
to various partial formulations that take into account only this or that facet of the problem. In this respect the counterion-only or the one-component Coulomb fluid model 
system has proved to be of substantial value \cite{hoda}. Heuristically as well as numerically.  A proper understanding of the behavior of charged systems would thus start with 
 the analysis of counter-ion distribution around charged macromolecular surfaces, neglecting completely the effects of salt.

Both the weak and the strong coupling approximations are based on a functional integral or field-theoretic representation \cite{podgornik} of the grand canonical partition
function of a system composed of fixed surface charges with intervening mobile counterions, and depend on the value of a single dimensionless coupling parameter $\Xi$ \cite{Netz}.
The distance at which two unit charges interact with thermal energy $k_{\mathrm{B}}T$ is known as the Bjerrum length $\ell_{\mathrm{B}}=e_0^2/(4\pi\varepsilon\varepsilon_0 k_{\mathrm{B}}T)$ 
(in water at room temperature, one has $\ell_{\mathrm{B}}\simeq 0.7$nm). If the charge valency of the counterions is $q$ then the aforementioned distance scales 
as $q^2 \ell_{\mathrm{B}}$. Similarly, the distance at which a counterion interacts with a macromolecular surface (of surface charge density $\sigma$) with 
an energy equal to $k_{\mathrm{B}}T$ is called the Gouy-Chapman length, defined as  $\mu=e_0/(2\pi q\ell_{\mathrm{B}}\sigma)$.
A competition between ion-ion and ion-surface interactions can be quantitatively measured with a ratio of these characteristic lengths, that is
 $\Xi=q^2 \ell_{\mathrm{B}}/\mu=2\pi q^3 \ell_{\mathrm{B}}^2\sigma/e_0$, which is known as the (Netz-Moreira) electrostatic coupling parameter \cite{Netz}.
The weak coupling (WC) regime ${\Xi}\ll 1$ (appropriate for low valency counterions and/or weakly charged surfaces), 
is characterized by the fact that the width of the counterion layer $\mu$ is much larger than the separation between two neighboring counterions in solution 
and thus the counterion layer behaves basically as a three-dimensional gas. Each counterion in this case interacts with many others and the collective
 mean-field approach of the Poisson-Boltzmann (PB) type is completely justified. On the other hand in the strong coupling (SC) regime ${\Xi}\gg 1$ (appropriate for high valency counterions and/or highly charged surfaces), the mean distance between counterions, $a_\bot \simeq \sqrt{qe_0/\sigma}$, is much larger than the 
layer width ({\em i.e.}, $a_\bot/\mu\sim \sqrt{\Xi}\gg 1$), indicating that the counterions are highly localized laterally and form a strongly correlated quasi-two-dimensional layer
next to a charged surface. In this case, the weak-coupling approach breaks down 
due to strong counterion-surface and counterion-counterion correlations. Since counterions can move almost independently from the others along the 
direction perpendicular to the surface, the collective many-body effects that enable a mean-field description are absent, necessitating a complementary SC description \cite{Netz}. 
The range of validity of both limiting theories at intermediate values of the coupling parameter
 has been explored thoroughly in the literature \cite{hoda,Naji,Netz,yoram,santangelo,lue,weeks}. 

Formally the weak coupling limit can be straightforwardly identified with the saddle-point approximation of the field theoretic representation of the grand 
canonical partition function, 
and is reduced to the mean-field PB theory in the lowest order for $\Xi\rightarrow 0$. The quadratic fluctuations 
around the mean field  provide a second-order correction to the mean-field solution for small finite $\Xi<1$ \cite{podgornik,orland,ramin,attard,safran-2,ha-2,lau-2}. The strong coupling approximation has no PB-like 
correlates \cite{Netz} since it is formally equivalent to a single particle description obtained from a systematic $1/\Xi$ expansion in the limit $\Xi\rightarrow \infty$, and corresponds to two lowest order terms in the virial expansion of 
the grand  canonical partition function. The consequences and the formalism of these two limits of the Coulomb fluid description have been explored widely 
and in detail (for reviews, see Refs. \cite{Naji,hoda}).

Considering the inhomogeneity of charged surfaces in various biological contexts it has always been of interest to investigate not just electrostatic interactions 
between symmetrical charged surfaces, {\em i.e.} those bearing equal charges of the same sign, but also interactions between surfaces bearing unequal charges or even
 charges of opposite sign \cite{adrian,benyaakov,lau1,safran,grunberg,sens,trulsson}. 
This problem has a venerable history starting from the seminal work of Parsegian and Gingell \cite{adrian} who formulated a linearized PB 
theory of the interactions in the presence of salt. The linearization {\em ansatz} was later generalized in the work of Lau and Pincus \cite{lau1} and 
Ben-Yaakov {\em et al.} \cite{benyaakov} who formulated the appropriate non-linear mean-field theory of non-symmetric electrostatic interactions. 

It is thus our goal in this contribution to show how and to what extent the asymmetry in the distribution of charges on two apposed planar surfaces affects the interactions between macromolecular surfaces carrying them. Below we shall present a complete analysis of the asymmetric case in the weak coupling limit, {\em i.e.} the mean-field Poisson-Boltzmann theory supplemented with a
quadratic-fluctuations analysis, as well as in the strong coupling limit {\em via} the asymptotic  strong-coupling theory and evaluate how these analytical results compare
with extensive numerical simulations. We will show that in their respective regimes of validity ({\em i.e.} small/large couplings) both approximations present a very
accurate quantitative statistical 
description of the system. 

\section{Geometry}
\label{sec:geo}

In our model system we consider uniform surface charge distributions on two plane-parallel  surfaces (located at $z = \pm a$) given by the surface 
charge density of the form
\begin{equation}
\rho_0(\Av r)=\sigma_1\,\delta(z+a)+\sigma_2\,\delta(z-a).
\end{equation}
We may interchangeably use the half-separation $a$, or 
\begin{equation}
D=2a
\end{equation}
to identify the surface-surface distance. 

\begin{figure}[t]
\centerline{\psfig{figure=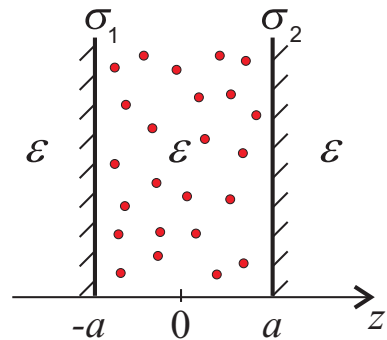,width=5cm}}
\caption{(Color online) Geometry of the system comprising two asymmetrically charged planar surfaces located at $z=\pm a$ (at separation distance $D=2a$)
 with neutralizing point-like counterions of valency $q$ distributed in between \cite{Note_2a}. }
\label{fig:geometry}
\end{figure}

We assume furthermore that the charge of both bounding surfaces is compensated by mobile counterions of charge valency $q$
immersed in an aqueous medium of dielectric constant $\varepsilon$ and distributed in between
the two surfaces (see Fig. \ref{fig:geometry}). We thus neglect all coions. This approximation is relevant for low salt concentrations 
where the Debye screening length is much larger than the scales of interest  \cite{olli}. 
We consider the surfaces as impenetrable to counterions and neglect the  dielectric discontinuity 
across the bounding surfaces which was addressed at various levels of approximation in \cite{Jho1,kanduc,jho-prl,discont}.

Without loss of generality we can assume here that $q>0$ and
\begin{equation}
\quad \sigma_1+\sigma_2<0,\quad{\textrm{and}}\quad \sigma_2>\sigma_1, 
\quad{\textrm{so that}}\quad\sigma_1<0.
\end{equation}
It will be helpful for our later developments to introduce an {\em asymmetry parameter} $\zeta$ that will allow us to quantify the dissimilarity between the two bounding surfaces as
\begin{equation}
\zeta=\frac{\sigma_2}{\sigma_1}>-1.
\end{equation}
Furthermore, by suitably normalizing the results one can concentrate exclusively on the interval $-1 \le \zeta \le 1$. All other cases can be 
mapped onto this interval with appropriate rescaling of the parameters. The values $\zeta = 1$ and $\zeta = -1$ represent exceptional points 
in the parameter space: $\zeta = 1$ is the standard symmetric case ($\sigma_1=\sigma_2$) already amply treated in the literature, 
and $\zeta = -1$ represents the antisymmetric case ($\sigma_1=-\sigma_2$) with no counterions between surfaces that reduces to the trivial
case of a planar capacitor. These two well-understood limiting cases will be thus omitted from our discussion.

Counterions between the surfaces satisfy electroneutrality condition that can be written in the form
\begin{equation}
Ne_0q+(\sigma_1+\sigma_2)S=0,  
\label{eq:electroneutral}
\end{equation}
where $S$ denotes the (infinite) area of each surface.  

\section{Dimensionless representation}
\label{sec:rescal}

Because of the asymmetry present in the system, we have two length scales describing the interaction of the counterions with each of the
 bounding surfaces. These two length scales are given by the corresponding Gouy-Chapman lengths associated with the two surfaces  as
\begin{equation}
\mu_1=\frac{e_0}{2\pi \ell_{\mathrm{B}} q |\sigma_1|}\equiv\mu,
\qquad
\mu_2=\frac{e_0}{2\pi \ell_{\mathrm{B}} q |\sigma_2|}=\frac{\mu}{\zeta}.
\end{equation}
For the same reason we can thus define two different coupling parameters
\begin{equation}
\Xi_1=\frac{q^2 \ell_{\mathrm{B}}}{\mu_1}\equiv\Xi,
\qquad
\Xi_2=\frac{q^2 \ell_{\mathrm{B}}}{\mu_2}=\zeta\Xi,
\end{equation}
each one being defined by the ratio between the Bjerrum length and the corresponding Gouy-Chapman length.
In what follows, we  rescale the surface separation as 
\begin{equation}
  \tilde D = D/\mu
\end{equation}
 (or the rescaled half-distance as $\tilde a = a/\mu$) with respect to plate 1. With an appropriate rescaling one could also equivalently 
define all dimensionless lengths with respect to plate 2.
Thus the minimal set of dimensionless parameters that fully characterize the 
system in the thermodynamic limit is given by
$\{\Xi, \zeta, \tilde D \}$. 
 
Other physical quantities such as the mean electrostatic potential $\psi(z)$, 
number density of counterions, $n(z)$,
and the pressure, $p$, acting on each surface can be rescaled as well.  We shall use
the standard rescaled electrostatic potential $$\tilde \psi(z) = \beta q e_0 \psi(z),$$as well as the rescaled density and pressure 
\begin{equation}
\tilde n(z) = \frac{n(z)}{2\pi\ell_{\mathrm{B}} (\sigma_1/e_0)^2}\quad{\textrm{and}}\quad
	\tilde p = \frac{\beta p}{2\pi\ell_{\mathrm{B}} (\sigma_1/e_0)^2},
\end{equation}
where $\beta=1/k_{\mathrm{B}}T$, and all the other quantities have been defined above.

\section{Mean-field Poisson-Boltzmann (PB) approximation}

In the weak coupling regime, the leading contribution to the partition function comes from the saddle-point configuration of the local fluctuating
 electrostatic potential, $\psi_0(z)$ \cite{podgornik}. 
The saddle-point configuration can be straightforwardly translated into a solution of the PB equation 
and corresponds to an exact asymptotic result in the limit $\Xi\rightarrow 0$ \cite{Netz,orland}. For a system containing only counterions, the PB equation for the 
dimensionless potential, $\tilde  \psi_0(z)$, can be written in the standard form \cite{Andelman,DLVO}
\begin{equation}
\frac{\mathrm{d}^2\tilde \psi_0(z)}{\mathrm{d}z^2}=-(4\pi \ell_{\mathrm{B}} q^2)\lambda_0 e^{-\tilde\psi_0(z)},
\label{eq:PBeq}
\end{equation}
with boundary conditions
\begin{eqnarray}
\left.\frac{\mathrm{d}\tilde\psi_0}{\mathrm{d}z}\right\vert_{-a}&=&\frac{2}{\mu},\nonumber\\
\left.\frac{\mathrm{d}\tilde\psi_0}{\mathrm{d}z}\right\vert_{a\phantom{-}}&=&-\frac{2\zeta}{\mu}.
\label{BC-rep}
\end{eqnarray}
Integration of the PB equation gives rise to the first integral of the system of the form
\begin{equation}
\beta p_0 = -\frac{1}{8\pi \ell_{\mathrm{B}} q^2}\bigg(\frac{\mathrm{d}\tilde\psi_0}{\mathrm{d}z}\bigg)^2 + n_0(z),
\label{rep-pressure}
\end{equation}
where the constant $p_0$ is nothing but the mean-field PB pressure acting between the bounding surfaces  \cite{Andelman} and
\begin{equation}
 n_0(z) = \lambda_0\,e^{-\tilde\psi_0(z)},
\end{equation}
 is the PB number density profile of counterions between the surfaces.
The normalization factor $\lambda_0$ follows from the electroneutrality condition (\ref{eq:electroneutral}) as
\begin{equation}
	\lambda_0 = -\frac{\sigma_1+\sigma_2}{qe_0\int_{-a}^{a} {\mathrm{d}}z\,e^{-\tilde\psi_0(z)} }. 
\end{equation}

The nature of the solution $\tilde\psi_0(z)$ obviously crucially depends on the sign of the pressure $p_0$ \cite{lau1,benyaakov}.
Different forms are obtained for positive and negative pressures, corresponding to repulsion and attraction between the bounding surfaces respectively.  We review these different cases separately.

\subsection{Repulsion regime $p_0>0$}

In the case of repulsive pressure the appropriate solution of Eq. (\ref{eq:PBeq}) can be written as
\begin{equation}
\tilde\psi_0=\ln\bigg\{\frac{\lambda_0}{\beta p_0}\,\cos^2\alpha(z-z_0)\bigg\},
\label{psi0}
\end{equation}
where the constants $z_0$ and $\alpha^2\equiv 2\pi \ell_{\mathrm{B}} q^2 (\beta p_0)$ are obtained
 from the boundary conditions (\ref{BC-rep}) and satisfy the set of two equations
\begin{eqnarray}
\alpha\,\tg\,\alpha(a+z_0)&=&\frac{1}{\mu},\label{z01}\\
\alpha\,\tg\,\alpha(a-z_0)&=&\frac{\zeta}{\mu}.\label{z02}
\end{eqnarray}
Eliminating $z_0$ we obtain an equation for $\alpha$ of the form
\begin{equation}
\tg\,(2\alpha a)=\frac{\alpha(\zeta+1)\mu}{\alpha^2\mu^2-\zeta}.
\end{equation}
The solution of this equation provides the final result for the repulsive PB pressure $p_0$.
Note in particular that in rescaled units and by definition one has
\begin{equation}
\tilde p_0=\tilde \alpha^2, 
\label{pres-pb-rep}
\end{equation}
where $\tilde \alpha = \alpha \mu$. 
Once $\alpha$ is known, the parameter $z_0$ can be simply obtained from Eqs. (\ref{z01}) or (\ref{z02}) and thus the potential $\tilde \psi_0(z)$, Eq. (\ref{psi0}), is  fully determined.
The density profile of counterions  then follows from 
\begin{equation}
n_0(z)=\frac{\beta p_0}{\cos^2\alpha(z-z_0)}. 
\label{densWC1}
\end{equation}

A positive (repulsive) solution for the pressure as considered in this section
is always possible for any given asymmetry parameter
$\zeta$ (excluding the trivial case of $\zeta=-1$). In particular,
it easily follows that within the mean-field theory 
two surfaces of equal sign ($\zeta>0$) always repel, that is at all separation distances $\tilde D$.  
When the surfaces bear charges of opposite sign $\zeta< 0$, they 
attract at large separations (see below) and a repulsion emerges only 
at sufficiently small separations.

At small separations $\tilde D\ll 1$, we can obtain the limiting solution for $\alpha$ and 
thus the limiting small-distance pressure $p_0$ as 
\begin{equation}
\tilde p_0(\tilde D)\simeq \frac{1+\zeta}{\tilde D}
\label{rep-limit1}
\end{equation}
for arbitrary $|\zeta|<1$ as noted in Scetion \ref{sec:geo}. This is of course nothing but the ideal-gas osmotic pressure of counterion
confined between the two plates ({\em i.e.}, $p_0=N k_{\mathrm{B}}T/(SD)$
in actual units),
which dominates over the energetic contributions at small separations.
At large separations $\tilde D\gg 1$ and for $\zeta>0$, we obtain the asymptotic expansion
\begin{equation}
\tilde p_0(\tilde D)=\frac{\pi^2}{\tilde D^2}\bigg[1-\cfrac{2(\zeta+1)}{\zeta\tilde D}+{\mathcal O}\bigg(\frac{1}{\tilde D^2}\bigg)\bigg],
\label{rep-limit-zeta}
\end{equation}
which is valid for $\tilde D \gg 2(1+1/\zeta)$. Thus in the limit $\tilde D \rightarrow \infty$,
the pressure behaves as
\begin{equation}
\tilde p_0(\tilde D)\simeq \frac{\pi^2}{\tilde D^2},
\label{rep-limit}
\end{equation}
which agrees with the asymptotic pressure between two equally charged surfaces ($\zeta=1$). 
For smaller $\zeta$ than determined above, {\em i.e.} for $\zeta^{-1}\gg\tilde D\gg 1$, one needs to invoke a different asymptotic expansion
and specifically for $\zeta\simeq 0$ (one surface being neutral), one obtains
\begin{equation}
\tilde p_0(\tilde D)\simeq \frac{\pi^2}{4\tilde D^2}.
\end{equation}
This latter asymptotic result may be obtained from the one in Eq. (\ref{rep-limit})
by redefining $\tilde D \rightarrow 2 \tilde D$. This
may be understood simply by noting that because of
 symmetry a system with $\zeta>0$ may be decomposed into 
two halves each with an effective asymmetry parameter $\zeta=0$. 

\subsection{Attraction regime $p_0<0$}

An attractive pressure on the mean-field level is possible only if the surfaces are oppositely charged $\zeta< 0$.
The appropriate solution in this case is given by
\begin{equation}
\tilde\psi_0=\ln\bigg\{\frac{\lambda_0}{\beta |p_0|}\,\sh^2\alpha(z-z_0)\bigg\},
\end{equation}
where the constants $z_0$ and $\alpha^2\equiv 2\pi \ell_{\mathrm{B}} q^2 (\beta |p_0|)$
can again be obtained from boundary conditions, this time in the form
\begin{eqnarray}
\alpha\,\cth\,\alpha(a+z_0)&=&-\frac{1}{\mu},\\
\alpha\,\cth\,\alpha(a-z_0)&=&-\frac{\zeta}{\mu}.
\end{eqnarray}
Eliminating $z_0$ we obtain an equation for $\alpha$ as
\begin{equation}
\cth\,(2\alpha a)=-\frac{\zeta+\mu^2\alpha^2}{\mu\alpha(1+\zeta)}.
\end{equation}
In this case we have in rescaled units 
\begin{equation}
\tilde p_0=-\tilde \alpha^2, 
\label{pres-pb-att}
\end{equation}
and for the density profile of counterions 
\begin{equation}
n_0(z)=\frac{\beta |p_0|}{\sh^2\alpha(z-z_0)}.
\label{densWC2}
\end{equation}

The asymptotic form of the attractive pressure at large separations $\tilde D \gg 1$
can be derived as
\begin{equation}
\tilde p_0(\tilde D)\simeq -\zeta^2\left(1-4\,\frac{1+\zeta}{1-\zeta}\,e^{2\zeta \tilde D}\right),
\label{att-limit}
\end{equation}
where $\zeta< 0$ as noted above.
For infinite separations, this pressure does {\em not} vanish and exponentially approaches $ -\zeta^2$ since for $\zeta < 0$
the system behaves partially as a simple capacitor.

\subsection{Zero pressure $p_0=0$}

In the case of charged surfaces with opposite sign ($\zeta< 0$), the large-distance attraction
regime and the short-distance repulsion regime merge at the point of zero pressure, $\tilde D = \tilde D^\ast$,
where the surfaces are at equilibrium.  In this case, the PB solution for the potential
reads 
\begin{equation}
\tilde\psi_0=\ln\bigg\{2\pi \ell_{\mathrm{B}} q^2 \lambda_0\,(z-z_0)^2\bigg\}, 
\end{equation}
and the density profile of counterions is given by
\begin{equation}
n_0(z)=\frac{1}{2\pi \ell_{\mathrm{B}} q^2\,(z-z_0)^2},  
\label{densWC3}
\end{equation}
where $z_0$ is found from the boundary conditions as $z_0=-\mu-D^\ast/2$, and the 
{\em bound-state separation}, $D^\ast$, follows in rescaled units as 
\begin{equation}
   \tilde{D}^\ast=-\frac{1+\zeta}{\zeta}.
\label{eq:PB_a*}
\end{equation}
The surfaces attract for $\tilde D >\tilde{D}^\ast$ and repel for $\tilde D < \tilde{D}^\ast$. 
In the vicinity of $\tilde{D}^\ast$, that is for $|\tilde D^\ast-\tilde D|\ll 1$, the pressure behaves as 
\begin{equation}
\tilde p_0\simeq  \frac{3\zeta^3}{1+\zeta^3}\,|\tilde D^\ast-\tilde D|. 
\end{equation}
This concludes the calculation of the inter-surface pressure on the mean-field PB level strictly valid for 
$\Xi\rightarrow 0$.

The preceding results may be summarized in a phase diagram shown in Fig. \ref{fig:bound-SC} in terms of $D^\ast$ and the asymmetry parameter $\zeta$
displaying the mean-field attraction and repulsion regimes separated by the boundary line (\ref{eq:PB_a*}).   The forms of the pressure here are completely consistent with those derived 
by Lau and Pincus \cite{lau1} {\em via} a different route.

\section{Weak-coupling (WC) analysis: Quadratic fluctuations around mean field}

The first non-zero correction to the saddle point is second order in the fluctuations of the local electrostatic potential around the mean-field PB solution, $\psi_0$. Our goal here is to calculate the corrections in pressure, $p_2(D)$, stemming from these quadratic fluctuations, which leads then to the total WC pressure
\begin{equation}
	p(D) = p_0(D)+p_2(D).
\end{equation}
This approach has correlates in many diverse areas of physics where fluctuations around a mean-field solution are important \cite{addrefpod} and goes under different names, though the physics is always the same. We may conventionally refer to the mean-field PB term, $p_0(D)$, and the fluctuations contribution, $p_2(D)$, as the {\em zeroth-order} and the {\em second-order} correction terms
on the WC level, respectively. This procedure formally also corresponds  to
a series expansion in powers of $\Xi$ (loop expansion) around the asymptotic mean-field solution
($\Xi\rightarrow 0$)  \cite{kanduc,podgornik,Netz,orland} 
and is thus expected to be valid for sufficiently small coupling parameters as will be determined later.
Note also that in this latter sense the second-order pressure turns out to
be proportional to $\Xi$, that is $p_2\sim \Xi$, and thus corresponds to
a first-loop correction \cite{Netz,orland}. 

In order to proceed, one needs to evaluate the appropriate Hessian of the field action in the
partition function and study its fluctuation spectrum (see Refs. \cite{podgornik,kanduc} for more details).
The Hessian of the field action can be derived in the form
\begin{equation}
H(\Av r,\Av r')=u^{-1}(\Av r,\Av r')+ \beta(e_0 q)^2 n_0(z)\,\delta^3(\Av r-\Av r'),
\end{equation}
where $u^{-1}(\Av r,\Av r')=-\varepsilon\varepsilon_0\nabla^2_{\Av r}\,\delta^3(\Av r-\Av r')$
is the inverse Coulomb operator and   $n_0(z)$
is the zeroth-order PB density as derived in the previous section. Hence,
\begin{equation}
(4\pi\ell_\mathrm{B}q^2)n_0(z)= \left\{
	\begin{array}{ll}
	 \cfrac{2\alpha^2}{\cos^2\alpha (z-z_0)} & \quad p_0 > 0,\\
	\\
	 \cfrac{2}{(z-z_0)^2} & \quad p_0 = 0,\\
	\\
	 \cfrac{2\alpha^2}{\sh^2\alpha (z-z_0)} & \quad p_0 < 0.
	\end{array}
\right.
\label{sign-eigen}
\end{equation}

The corresponding correction, ${\mathcal F}_2$,  to the free energy of the system
is then given by the trace-log of the Hessian. It can be written equivalently in the following form \cite{podgornik,attard}
\begin{equation}
  \beta {\mathcal F}_2 = \frac{1}{2}\Tr\,\ln\,H(\Av r,\Av r')
= \frac{S}{4\pi}\int_0^{\infty} Q\>\ln\frac{{\mathcal D}_1( Q)}{{\mathcal D}_0(Q)}\,{\mathrm{d}}Q.
\label{trace}
\end{equation}
This form can be derived rather straightforwardly by using  the argument principle \cite{attard} and converting the discrete sum of eigenvalues of the
 Hessian operator  into an integral over the transverse wave-vector $\Av Q = (Q_{x}, Q_{y})$, with density  of modes $S/(2\pi)^2$ of the logarithm of
the secular determinant ${\mathcal D}_\lambda$ of the same operator.  The index $\lambda$ in the secular determinant  refers to the eigenvalue equation  that can be derived in the form
\begin{equation}
\Bigl(\frac{
\partial^2}{
\partial z^2}-Q^2-\lambda(4\pi\ell_\mathrm{B}q^2)n_0(z)\Bigr) f_{\lambda}(\Av Q,z)=0.
\label{eigF}
\end{equation}
By simply writing ${\mathcal D}(a, Q)$ for the quotient  ${{\mathcal D}_1( Q)}/{{\mathcal D}_0(Q)}$, and noting that the secular determinant depends
explicitly also on the value of the inter-surface spacing, $a$, the free energy contribution
from the quadratic fluctuations can be equivalently expressed exactly in a dimensionless form as 
\begin{equation}
\frac{\tilde{\mathcal F}_2}{\tilde{S}}=\frac{1}{2}{\Xi}\int_0^\infty \tilde{Q}\,\ln\,{\mathcal D}(\tilde{a}, \tilde{Q})\, {\mathrm{d}}\tilde{Q}.  
\label{second-f}
\end{equation}
where  $\tilde{{\mathcal F}}_2 = 2\varepsilon\varepsilon_0{\mathcal F}_2/\sigma_1^2\mu^3$,  $\tilde{Q} = \mu Q$ and the rescaled area $\tilde S=S/\mu^2$. 
Here the secular determinant of the Hessian for homogeneous transverse modes has been written as a function of dimensionless quantities
 $\tilde{a} = a/ \mu$, $\tilde{Q} = \mu Q$. This determinant has to be standardly regularized so that all irrelevant constants, {\em i.e.}
 all the terms not depending on the separation between the bounding surfaces, are dropped, amounting to a rescaling
\begin{equation}
{\mathcal D}(\tilde{a}, \tilde{Q})\rightarrow \frac{{\mathcal D}(\tilde{a}, \tilde{Q})}{{\mathcal D}(\tilde{a} \rightarrow \infty, \tilde{Q})}.
\label{renorm}
\end{equation}
This corresponds to a subtraction of the part of the free energy for two separate interfaces at infinite separation from the total free energy.

In the next step one has to calculate the secular determinant ${\mathcal D}(\tilde{a}, \tilde{Q})$ for each of the pressure regimes separately, since 
the appropriate eigenfunctions of the Hessian depend on the mean-field solution that in its turn depends on the sign of the interaction pressure, 
see Eq. (\ref{sign-eigen}). In what follows we shall follow closely the derivations in Refs. \cite{kanduc,attard,podgornik}.

The total pressure in the weak-coupling limit is thus the sum of the PB pressure and the
quadratic fluctuations correction and can be written as
\begin{equation}
\tilde p(\tilde D) = \tilde p_0(\tilde D) + \tilde p_2(\tilde D) = \tilde p_0(\tilde D)
		- \frac{1}{\tilde S}\left(\frac{\partial \tilde {\mathcal F}_2}{ \partial \tilde D}\right).
\label{secondord}
\end{equation}

\subsection{Repulsion regime $p_0>0$}

In this regime the secular determinant of the Hessian, Eq. (\ref{eigF}), can be obtained by solving
\begin{equation}
\left(\frac{{\mathrm{d}}^2}{{\mathrm{d}}z^2}-Q^2-\frac{2\alpha^2}{\cos^2\alpha(z-z_0)}\right)y(Q, z)=0,
\label{sec-rep}
\end{equation}
with appropriate boundary conditions implying continuity of the solution and its derivative across the bounding surfaces at $z = \pm a$. 

The general solution of Eq. (\ref{sec-rep}) for various regions in the perpendicular direction  can be written in the form
\begin{equation}
y(Q, z) = \left\{
   \begin{array}{ll}
	Ae^{Qz}  & \quad z<-a,\\
	\\	
  	B y_1+Cy_2 & \quad -a<z<a,\\
	\\
	De^{-Qz} & \quad z>a, 
	\end{array}
	\right.
\end{equation}
where 
\begin{eqnarray}
	&&y_{1}=e^{Qz}\left[1+\cfrac{\alpha}{Q}\,\tg\,\alpha(z-z_0)\right],\\
	&&y_{2}=\cfrac{Q^2}{Q^2+\alpha^2}\>e^{-Qz}\left[1- \cfrac{\alpha}{Q}\,\tg\,\alpha (z-z_0)\right].
\end{eqnarray}
Taking into account the continuity of the solution and its derivatives we get a set of four homogeneous equations for the coefficients $A, B, C$ and $D$. 
The solution exists only if the (secular) determinant of this system equals zero. Thus we derive the secular determinant of the Hessian operator in this case in the rescaled form
\begin{equation}
{\mathcal D}(\tilde a, \tilde Q)=\frac{\Gamma_+(\tilde \alpha,\tilde Q)-\Gamma_+(\tilde \alpha,0)e^{-4\tilde Q\tilde a}}{\tilde Q^2+\tilde \alpha^2},
\label{eq:secularD_final}
\end{equation}
where
\begin{equation}
\Gamma_+(\tilde\alpha,\tilde Q)=(1+\tilde\alpha^2+2\tilde Q+2\tilde Q^2)(\zeta^2+\tilde\alpha^2+2\zeta \tilde Q+2\tilde Q^2).
\end{equation}
The regularized form of the secular determinant is obtained by taking the quotient as indicated in Eq. (\ref{renorm}). While doing this,
 it is important to realize that $\tilde\alpha$ also depends on the inter-surface distance. In fact from Eq. (\ref{rep-limit}) it follows
 that the appropriate limit of $\tilde\alpha$ is
\begin{equation}
\lim_{\tilde a\rightarrow \infty}\tilde\alpha(\tilde a) = 0.
\end{equation}
In the regularization of the secular determinant this limiting behavior should be consistently taken into account.

Finally, the dimensionless quadratic fluctuations free energy $\tilde {\mathcal F}_2$
can be calculated numerically {\em via} Eqs. (\ref{second-f}) and (\ref{eq:secularD_final}).
The fluctuations contribution to the pressure, $\tilde p_2$, then follows from Eq. (\ref{secondord}).

The asymptotic form of the second-order dimensionless pressure can be obtained analytically.
Note that at large separations $\tilde D\rightarrow \infty$, a repulsive mean-field pressure $p_0>0$, as considered in this section, is possible only for non-negative $\zeta$. For not too small $\zeta > 0$, {\em i.e.} when $\tilde D\gg 2(1+1/\zeta)$, we find
\begin{equation}
\tilde p_2(\tilde D)\simeq-\Xi\,\pi^2\,\frac{\ln\, \tilde D}{\tilde D^3},
\end{equation}
while for $\zeta\simeq 0$, {\em i.e.} when $\zeta^{-1}\gg\tilde D\gg 1$, we get
\begin{equation}
\tilde p_2(\tilde D)\simeq-\Xi\,\pi^2\,\frac{\ln\, \tilde D}{8\tilde D^3}.
\end{equation}
Again the difference in the two cases above 
is due to the symmetry of the problem in the latter case, that can be described by redefining $\tilde D \rightarrow 2 \tilde D$
and discarding the sub-dominant terms.

The second-order pressure  is obviously {\em attractive} and in this regime leads to
a reduction of the total pressure from the mean-field value $p_0$. This clearly shows that
electrostatic correlations favor attraction between two repelling asymmetrically charged plates.
However, the total pressure never becomes  negative as the fluctuations
are assumed to be small within the second-order weak-coupling analysis.

\subsection{Attraction regime $p_0<0$}

In this case the secular determinant of the Hessian, Eq. (\ref{eigF}), is obtained by solving
\begin{equation}
\left(\frac{d^2}{dz^2}-Q^2-\frac{2\alpha^2}{\sh^2\alpha(z-z_0)}\right)y(Q, z)=0.
\end{equation}
The general solutions for particular regions in the $z$ direction are
\begin{equation}
y(Q, z) = \left\{
   \begin{array}{ll}
	Ee^{Qz}  & \quad z<-a,\\
	\\
  	F y_1+Gy_2 & \quad -a<z<a,\\
	\\
	He^{-Qz} & \quad z>a,
	\end{array}
	\right.
\end{equation}
where
\begin{eqnarray}
	&&y_{1}=e^{Qz}\left[1- \cfrac{\alpha}{Q}\,\cth\,\alpha (z-z_0)\right],\\
	&&y_{2}=\cfrac{Q^2}{Q^2-\alpha^2}\,e^{-Qz}\left[1+\cfrac{\alpha}{Q}\,\cth\,\alpha (z-z_0)\right].
\end{eqnarray}
Again the solution exists only if the determinant of the system of equations, which connect coefficients $E, F, G, H$ and stems from
 the application of the boundary conditions at $ z = \pm a$, is identically zero. This again defines the secular determinant 
${\mathcal D}(a, Q)$ appropriate for this case. It is easy to show that the secular determinant  ${\mathcal D}(a, Q)$ can be obtained from 
the $p_0>0$ result, Eq. (\ref{eq:secularD_final}),   simply by substituting  $\alpha^2\rightarrow-\alpha^2$ and so we can write
\begin{equation}
{\mathcal D}(\tilde a, \tilde Q)=\frac{\Gamma_-(\tilde\alpha,\tilde Q)-\Gamma_-(\tilde\alpha,0)e^{-4\tilde Q\tilde a}}{\tilde Q^2-\tilde\alpha^2},
\end{equation}
where
\begin{equation}
\Gamma_-(\tilde\alpha,\tilde Q)=(1-\tilde\alpha^2+2\tilde Q+2\tilde Q^2)(\zeta^2-\tilde\alpha^2+2\zeta \tilde Q+2\tilde Q^2).
\end{equation}
Here we can again regularize the secular determinant to discard divergences, Eq. (\ref{renorm}).
Again one has to be careful by taking the correct limit for $\tilde\alpha$ in the above regularization scheme. In this regime,
the appropriate limit is given by
\begin{equation}
\lim_{\tilde a\rightarrow \infty}\tilde\alpha(\tilde a) = - \zeta,
\end{equation}
as follows straightforwardly from Eq. (\ref{att-limit}).
The fluctuations contribution to the pressure, $\tilde p_2$,  can then be evaluated numerically  from Eq. (\ref{secondord}).

The asymptotic form of $\tilde p_2$ for $\tilde D\gg 1$ can be derived analytically as
\begin{equation}
\tilde p_2(\tilde D)\simeq \Xi\,f(\zeta)\,e^{2\zeta \tilde D},
\end{equation}
which is applicable only for charged surfaces of opposite sign, $\zeta<0$, which can attract ($p_0<0$)
at large separations. 
The function $f(\zeta)$ is defined as
\begin{equation}
f(\zeta)=\zeta^3\frac{1+\zeta}{1-\zeta}\left(\frac{2\arctan{\sqrt{1-2\zeta^2}}}{\sqrt{1-2\zeta^2}}+\ln\,\frac{1-\zeta^2}{2\zeta^2}\right).
\end{equation}
for $-\sqrt{2}/2<\zeta<0$, and
\begin{equation}
f(\zeta)=\zeta^3\frac{1+\zeta}{1-\zeta}\left(\frac{2\tanh^{-1}{\sqrt{2\zeta^2-1}}}{\sqrt{2\zeta^2-1}}+\ln\,\frac{1-\zeta^2}{2\zeta^2}\right).
\end{equation}
for $-1<\zeta<-\sqrt{2}/2$.
 The second-order pressure thus asymptotically  decays exponentially and can be only attractive. For not too small values of $\zeta$,
it is thus qualitatively very different from the case $\zeta \geq 0$.

\begin{figure}[!h]
\centerline{\psfig{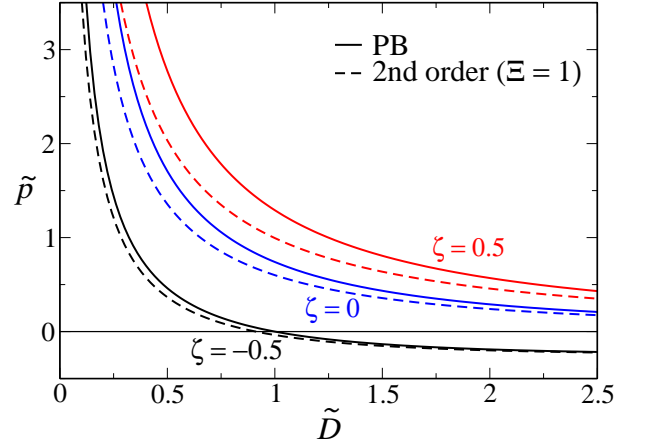}}
\caption{(Color online) Rescaled weak-coupling inter-surface pressure, Eqs. (\ref{pres-pb-rep}) and (\ref{pres-pb-att}), as a function of the rescaled distance, $\tilde D$,
 between two charged plates for three different asymmetry parameters $\zeta=-0.5, 0$ and 0.5 as shown on the graph.
The total pressure, $\tilde p=\tilde p_0+\tilde p_2$  (dashed lines, plotted here for $\Xi=1$ using Eq. (\ref{secondord})),
 is always lowered  from its mean-field
PB value, $p_0$ (solid lines, obtained for $\Xi\rightarrow 0$), since quadratic fluctuations around mean field
 favor attraction.
}
\label{fig:WC_p}
\end{figure}

The total weak-coupling pressure, $\tilde p=\tilde p_0+\tilde p_2$,
is shown in Fig. \ref{fig:WC_p} for a few different asymmetry parameters and a relatively small value of the coupling parameter $\Xi$.

\subsection{Regime of validity of the weak-coupling theory}
\label{sec:WC_crit}

As noted above the foregoing weak-coupling analysis is valid as long as the quadratic corrections are sufficiently
small so that the series expansion around the mean-field solution
does not diverge \cite{Netz,orland}.  As an approximate measure for the validity regime of this scheme, one can require that the second-order
correction term is smaller than the leading order term, {\em i.e.}
\begin{equation}
  |\tilde p_2| < |\tilde p_0|.
\end{equation}
This leads to a useful criterion identifying the regime of coupling parameters and distances in which the
weak-coupling theory is applicable.
For $p_0>0$ and by employing the closed-form expressions obtained for large separations $\tilde D \gg 1$, we find
the validity condition
\begin{equation}
     \Xi<\frac{\tilde D}{\ln\, \tilde D}.
\end{equation}
This indicates that at a given non-vanishing $\Xi$, the weak-coupling scheme becomes increasingly more
accurate at larger separations, while as
the surfaces get closer a smaller coupling parameter needs to be chosen.


On the other hand, for $p_0<0$ (which occurs for $\zeta<0$) and at large separations $\tilde D \gg 1$, we obtain
\begin{equation}
     \Xi<\frac{\zeta^2}{|f(\zeta)|}e^{-2\zeta \tilde D}.
     \label{eqlimit-wc}
\end{equation}
The right hand side here is exponentially large meaning that for charged surfaces of opposite sign, 
the weak-coupling analysis performs far better at finite coupling parameters and smaller
 inter-surface separations than for the surfaces of equal sign ($\zeta>0$).

Finally, note that for $p_0=0$ that corresponds to the equilibrium phase boundary line in Fig. \ref{fig:bound-SC} for $\zeta<0$, we deal
with a situation where the leading order term is zero and the fluctuations are {\em dominant} at any finite
value of $\Xi$. The convergence of the loop expansion has to be determined in this case by evaluating the higher order terms which
we shall not consider in this paper.

\section{Strong-coupling (SC) theory}

\begin{figure}[!h]
\centerline{\psfig{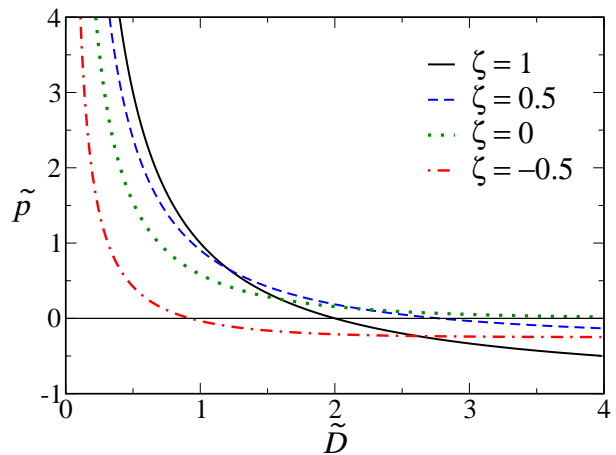}}
\caption{(Color online) Rescaled strong-coupling inter-surface pressure,  Eq. (\ref{eq:P_SC}), as a function of the rescaled distance, $\tilde D$,
 between two charged plates for different asymmetry parameters $\zeta=-0.5, 0, 0.5$ and 1 as shown on the graph.
Long-distance attractive pressure is suppressed for the $\zeta=0$ case. For symmetrically charged surface ($\zeta=1$),
we recover the standard SC result  $\tilde p(\tilde D) = -1+2/\tilde D$ \cite{Netz}. }
\label{fig:SC_p}
\end{figure}

The strong-coupling approximation coincides with the lowest order non-trivial expansion of the partition function in terms of the fugacities of
the counterions. This expansion may be expressed as a $1/\Xi$ series expansion \cite{Netz},
whose leading order term ($\Xi\rightarrow \infty$) corresponds to the so-called SC theory.
We will not delve into the strong-coupling expansion in more detail since it has
been exhaustively reviewed in the literature \cite{hoda,Naji,Netz}.
On the leading order, the free energy is obtained as
\begin{equation}
{\mathcal F}=W_0-Nk_\mathrm{B}T\,\ln\int e^{-\beta(W_1+W_2)}\mathrm{d}V,
\label{defF}
\end{equation}
where $W_0$ is electrostatic interaction energy of charged surfaces
\begin{equation}
W_0=\frac{\sigma_1\sigma_2}{2\varepsilon\varepsilon_0}\,S\,D,
\end{equation}
with $S$ representing surface area of each plate, and $W_1$ and $W_2$ are electrostatic interaction
energies between a single counterion
and individual charged surfaces, {\em i.e.}
\begin{equation}
W_1=-\frac{qe_0\sigma_1}{2\varepsilon\varepsilon_0}\big(\frac{D}{2}+z\big),
\qquad
W_2=-\frac{qe_0\sigma_2}{2\varepsilon\varepsilon_0}\big(\frac{D}{2}-z\big).
\end{equation}
Since in the strong-coupling regime the free energy is given {\em via} simple quadratures, it is much simpler to evaluate it
than on the weak-coupling level. Defining the rescaled free energy
\begin{equation}
\tilde{\mathcal F}=\frac{2\varepsilon\varepsilon_0}{\sigma_1^2\mu^3}{\mathcal F},
\end{equation}
we obtain
\begin{equation}
\frac{\tilde {\mathcal F}}{\tilde S}=(1+\zeta^2)\frac{\tilde D}{2} -(1+\zeta)\,\ln\,\sinh\bigg[(1-\zeta)\frac{\tilde D}{2}\bigg].
\end{equation}
Differentiating the free energy with respect to the surface-surface
distance $\tilde D $ we get the corresponding pressure acting between the bounding surfaces
\begin{equation}
\tilde p(\tilde D)=-\frac{1}{2}(1+\zeta^2)+\frac{1}{2}(1-\zeta^2)\,\coth\bigg[(1-\zeta)\frac{\tilde D}{2}\bigg].
\label{eq:P_SC}
\end{equation}
The dependence of this dimensionless pressure on the separation  for different values of $\zeta$ is presented in Fig. \ref{fig:SC_p}.
Note that the SC pressure can become attractive for {\em both} like-charged and oppositely charged surfaces which
contrasts with the mean-field theory that does not allow attraction between like-charged surfaces.
This is because of the strong electrostatic correlations mediated by counterions between the
charged surfaces for $\Xi\gg 1$ and has been investigated throughly before for equally charged surfaces \cite{Netz}.
Our results show that a similar attraction mechanism holds for asymmetrically charged surfaces in the SC limit.

The pressure exhibits two well-defined limiting laws obtainable for small and large inter-surface separations. 
For small separations $\tilde D\ll 1$, we have 
\begin{equation}
\tilde p(\tilde D)\simeq
	\cfrac{1+\zeta}{\tilde D}, 
\label{pSC_limit}
\end{equation}
for arbitrary $|\zeta|<1$ as noted in Section \ref{sec:geo}. For large separations $\tilde D\gg 1$, we obtain
\begin{equation}
\tilde p(\tilde D) \simeq-\zeta^2.
\label{newlimit}
\end{equation}
Note that in the limit $\tilde D\rightarrow 0$ the SC pressure coincides with the PB result on the leading order (compare Eqs. (\ref{rep-limit1}) and (\ref{pSC_limit})) and  represents the  ideal-gas osmotic pressure of counterions which dominates over the electrostatic contributions. The PB and SC forms for $\zeta<0$ coincide also in the limit of $\tilde D\gg 1$ (compare Eqs. (\ref{att-limit}) and (\ref{newlimit})) and reduce to the pressure in the capacitor.

\begin{figure}[!h]
\centerline{\psfig{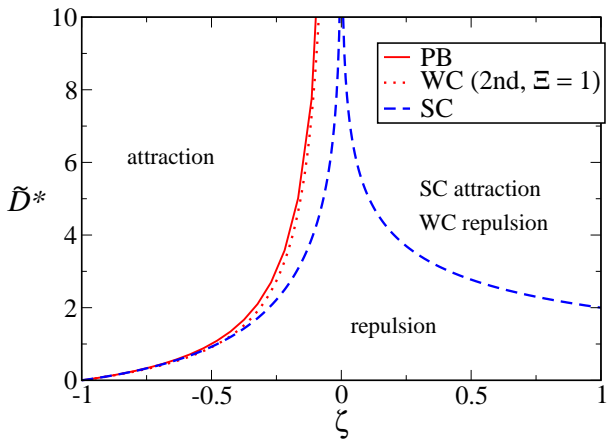}}
\caption{(Color online) Rescaled bound-state surface-surface separation  $\tilde D^\ast=2\tilde  a^\ast$ as a function of the asymmetry parameter $\zeta$ as predicted by  the zeroth-order PB theory (solid line), 
Eq. (\ref{eq:PB_a*}), the second-order WC theory at $\Xi=1$ (dotted line), and the  strong-coupling theory (dashed line), Eq. (\ref{eq:D*_SC}). 
Surfaces attract for $\tilde D>\tilde D^\ast$ and repel otherwise. Only in the SC limit can two surfaces of equal sign ($\zeta>0$)  attract. The  second-order WC result is obtained by finding the zero-pressure point of the total WC pressure $\tilde p(\tilde D)=\tilde p_0(\tilde D)+\tilde p_2(\tilde D)$, Eq. (\ref{secondord}), for given $\Xi$ and $\zeta$.}
\label{fig:bound-SC}
\end{figure}

The dependence of the
pressure on the inter-surface separation points to the existence of a bound state defined {\em via} $p(D^\ast)=0$.
The SC bound-state separation $D^\ast$ can be expressed analytically as
\begin{equation}
\tilde D^\ast=-\frac{2\,\ln\vert\zeta\vert}{1-\zeta},
\label{eq:D*_SC}
\end{equation}
and is presented in Fig. \ref{fig:bound-SC} as a function of $\zeta$. Obviously both like-charged and oppositely charged surfaces can form bound states at small surface-surface separations.
The bound-state separation approaches infinity and the surfaces  {\em unbind} asymptotically as  $\zeta \rightarrow 0$, that is when one plate becomes electroneutral. 

Finally one can also derive the explicit form of the counterion density as a function of the normal coordinate $z$. This can be read off 
simply from the integrand in Eq. (\ref{defF}), 
that is  $n(z) = C \exp(-\beta(W_1+W_2))$, where $C$ is a normalization 
factor \cite{Netz,Naji}. According to the electroneutrality condition, we normalize the density as $\int_{-a}^a n(z)\,{\mathrm{d}}z = -(\sigma_1+\sigma_2)/qe_0$, or in rescaled units
\begin{equation}
\int_{-\tilde a}^{\tilde a} \tilde n(\tilde z)\,{\mathrm{d}}\tilde z=1+\zeta,
\end{equation}
where we have defined $\tilde z=z/\mu$ with $\mu=-e_0/(2\pi \ell_{\mathrm{B}} q\sigma_1)$ being
the Gouy-Chapman length with respect to plate 1. 
From here, the density profile is obtained as
\begin{equation}
\tilde n(z)=\frac{1-\zeta^2}{2}\,\frac{e^{-(1-\zeta)\tilde z}}{\sinh\big[(1-\zeta)\tilde D/2\big]},
\label{densSC}
\end{equation}
as a function of $z$ and $\zeta$.

\subsection{Regime of validity of the strong-coupling theory}
\label{sec:SC_crit}

The regime of applicability of the leading order SC theory follows from a simple criterion
that has been discussed and confirmed previously in the case of equally charged
surfaces by both MC simulations and higher-order calculations \cite{hoda,Naji,Netz}. The generalization
to asymmetrically charged surfaces is straightforward.

For large couplings,  counterions are strongly attracted to an oppositely charged surface as the counterion-surface
interaction becomes large and equivalently, the Gouy-Chapman length, $\mu$, 
 and the thickness of the counterionic layer at the surface become small. The layer thickness has to be compared
with the typical lateral spacing between  counterions $a_\bot$. 
For counterions sandwiched between two asymmetrically charged surfaces, this latter 
quantity follows from the local electroneutrality condition as
\begin{equation}
   a_\bot^2 \simeq -\frac{qe_0}{\sigma_1+\sigma_2},
\end{equation}
up to a factor of the order unity  and assuming that the surfaces are sufficiently close so that they
may be strongly coupled {\em via} the counterions as will
be determined consistently here.
In rescaled units, one gets
\begin{equation}
   \tilde a_\bot^2 \simeq \frac{\Xi}{1+\zeta},
\end{equation}
where $\tilde a_\bot = a_\bot/\mu$.
Obviously,  $\tilde a_\bot$ becomes large relative to the layer thickness as $\Xi$ grows.
Note that on the other hand the lateral Coulomb repulsion between counterions in this quasi-two-dimensional
layer becomes much larger than the thermal energy, {\em i.e.} $q^2\ell_{\mathrm{B}}/a_\bot \sim \sqrt{\Xi} \gg 1$,
indicating that counterions form a strongly correlated liquid in which they are highly localized
within correlation holes of lateral size $\tilde a_\bot \sim \sqrt{\Xi} \gg 1$ \cite{hoda,Naji,Netz}. Thus
for surface-surface separations, $\tilde D$, smaller than the correlation hole size, {\em i.e.}
\begin{equation}
   \tilde D \ll \sqrt{\frac{\Xi}{1+\zeta}},
\end{equation}
counterions can move almost independently from each other in the  direction normal to the surface and 
one can safely assume that the effective surface-surface interaction as well as the counterionic density profile
follow only from the interactions of {\em individual} counterions with the bounding charged surfaces.
The counterion-counterion interactions contribute on the sub-leading order and matter at larger separations. 
This picture is of course confirmed on a systematic level by the SC expansion analysis \cite{hoda,Naji,Netz}, and
the above equation sets a criterion for the validity regime of the single-particle leading order  SC theory ($\Xi\rightarrow\infty$), 
when applied to finite coupling parameters.

\section{Simulations}

We performed Monte-Carlo (MC) simulations in order to study the system of two 
asymmetrically charged surfaces beyond the analytical limits
of weak and strong coupling discussed above.  
All simulations were performed in the Canonical ensemble (NVT) using the standard Metropolis algorithm \cite{Metropolis:53}.
The mobile counterions were modeled as point charges \cite{Note_2a} enclosed in a simulation box bounded in $z$
direction by two charged surfaces of distance $D$ and surface charge densities $\sigma_1$ and $\sigma_2$ (compare Fig. \ref{fig:geometry}).
Periodic boundary conditions were applied in the lateral directions parallel to the bounding surfaces. 
The lateral size of the charged surfaces, $L$, which is set equal to the lateral size of the simulation box, was held fixed throughout all the simulations.
The number of counterions, $N$, was varied between 60 and 1800, depending on the system parameters in
order to fulfill the electroneutrality condition. The counterions interact through the Coulombic potential
\begin{equation}
U(r_{ij}) = \frac{q_iq_j}{4 \pi \varepsilon\varepsilon_0 r_{ij}},
\end{equation}
with $q_i=e_0q$ being the charge of the $i$-th counterion and  $r_{ij}$ the separation distance between the $i$-th and
the $j$-th counterions (with $i, j=1,\ldots,N$). The interaction energy of the $i$-th counterion
 with the charged surfaces in the simulation box are given by
\begin{eqnarray}
U_k(z_{ik}) =
	\frac{q_i \sigma_k}{4 \pi \varepsilon\varepsilon_0}
		\bigg[4L\, \ln\frac{\sqrt{L^2/2+z_{ik}^2}+L/2}{\sqrt{(L/2)^2+z_{ik}^2}}
	\nonumber \\
	-2z_{ik}\bigg\{ \arcsin\bigg(\frac{(L/2)^4-z_{ik}^4 -(L^2/2)z_{ik}^2}{[(L/2)^2+z_{ik}^2]^2}\bigg)
		+ \frac{\pi}{2}\bigg\}\bigg],
\label{Coulomb_Wallterm}
\end{eqnarray}
where $\sigma_k=\sigma_1$ and $\sigma_2$ is surface charge density of the $k$-th surface (with $k=1,2$) and  $z_{ik}$ is the normal distance between the $i$-th counterion and the $k$-th surface.
The long-ranged Coulomb interactions in this system were accounted for
{\em via} a charged sheet scheme similar to that proposed by Torrie and Valleau \cite{Torrie:82}. This scheme makes use of the
counterion profile in the simulation box in order to calculate an external field, stemming from the long-ranged interactions.
This external field is iteratively updated and self-consistency is achieved normally in a few iterations.

In the course of simulations, new configurations were created by trial displacements of 
the counterions and equilibration was accomplished by running through
$10^6$ configurations. The $10^7$ following configurations were then used for the production runs.

The pressure $p$ was calculated in the production runs according to the contact-value theorem as
\begin{equation}
p_k=k_{\mathrm{B}} T n^{\textrm{contact}}_k - \frac{\sigma_k^2}{ 2 \varepsilon\varepsilon_0},
\end{equation}
where $n^\textrm{contact}_k$ is the density of counterions at contact with the $k$-th surface. In thermodynamic equilibrium, 
the pressure does not depend on which surface ($k =1$ or $k = 2$) is chosen in order to calculate the pressure from the 
above equation, and the contact condition at both surfaces leads to precisely the same value for the 
pressure. All simulations were conducted at fixed temperature $T=298$~K, lateral simulation box size $L=245$~\AA\,
and dielectric constant $\varepsilon=  78.7$, which is assumed to be the same throughout the system.

The simulations were performed at different values of the coupling parameter $\Xi$ and the asymmetry parameter $\zeta$. We 
 explored the $\Xi$ parameter space extensively by using $\Xi = 0.32, 0.64, 2.5, 3.2, 5.1, 6.4$, $8.6, 17, 25, 51, 86$ and $172$ in order 
to cover exhaustively both the weak coupling and the strong coupling regimes.
The concurrent values of the asymmetry parameter were always taken as $\zeta = -0.5,\> 0,\> +0.5$ at each value of the coupling parameter. 
The results are plotted in the form of rescaled density and pressure as previously defined in this paper
(Section \ref{sec:rescal}).




\section{Discussion}

In order to asses the validity of the weak and strong coupling results presented above 
for asymmetrically charged surfaces, we performed extensive MC simulations and compared them to analytical results in both limits. 
It transpires from this comparison that the simulation results corresponding to an exact evaluation of the partition function 
are always bracketed by the WC and the SC limiting forms, smoothly approaching them in the appropriate limits of the coupling parameter $\Xi$.

\begin{figure}[!h]
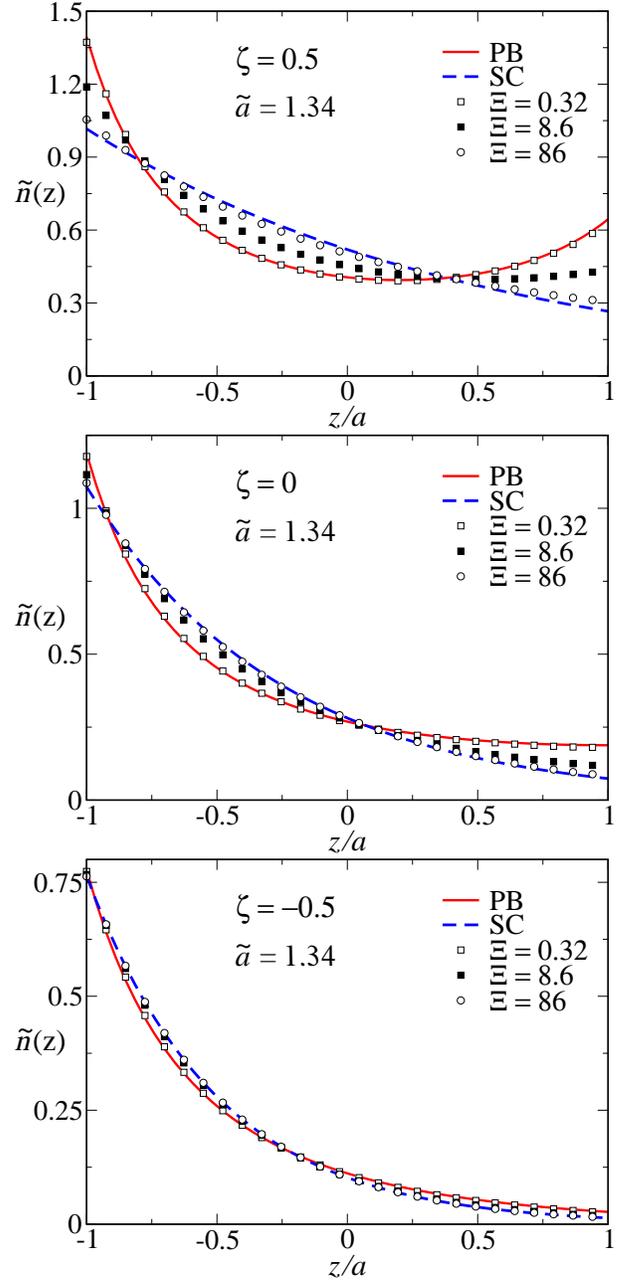

\centerline{\psfig{figure=dens1.eps,width=8cm}}
\centerline{\psfig{figure=dens2.eps,width=8cm}}
\centerline{\psfig{figure=dens4.eps,width=8cm}}
\caption{(Color online) Rescaled counterion density profile $\tilde n(z)$ between two asymmetrically charged surfaces at half-separation $\tilde a=\tilde D/2=1.34$ for three different
asymmetry parameters $\zeta=0.5, 0$ and $-0.5$ (top to bottom). Solid lines represent the PB prediction, Eqs. (\ref{densWC1}) and (\ref{densWC2}), and the dashed lines show the
SC prediction, Eq. (\ref{densSC}). Symbols correspond to MC simulations data at three different coupling parameters $\Xi=0.32$ (open squares), 8.6 (filled squares) and 86 (open circles).}
\label{dens}
\end{figure}

First we compare the density profiles of simulations with theoretical results given by 
Eqs. (\ref{densWC1}) and (\ref{densWC2}) for the PB limit (solid lines) and Eq. (\ref{densSC}) for the SC limit (dashed lines). 
As seen from Fig. \ref{dens} the theoretical PB and SC rescaled density profiles 
represent two extremal cases and all MC simulations results with finite values 
of $\Xi$ are located consistently between these two limits.
The MC results for small $\Xi$ are almost exactly spot on the PB prediction, while larger discrepancies are observed as $\Xi$ grows. 
For large enough  $\Xi>10$, the MC results slowly converge to the SC result. This is especially clear for surfaces with charges of 
equal sign, $\zeta=0.5$, whereas for surfaces with opposite sign, $\zeta=-0.5$, there is no big difference between 
PB and SC profiles.

\begin{figure}[!h]
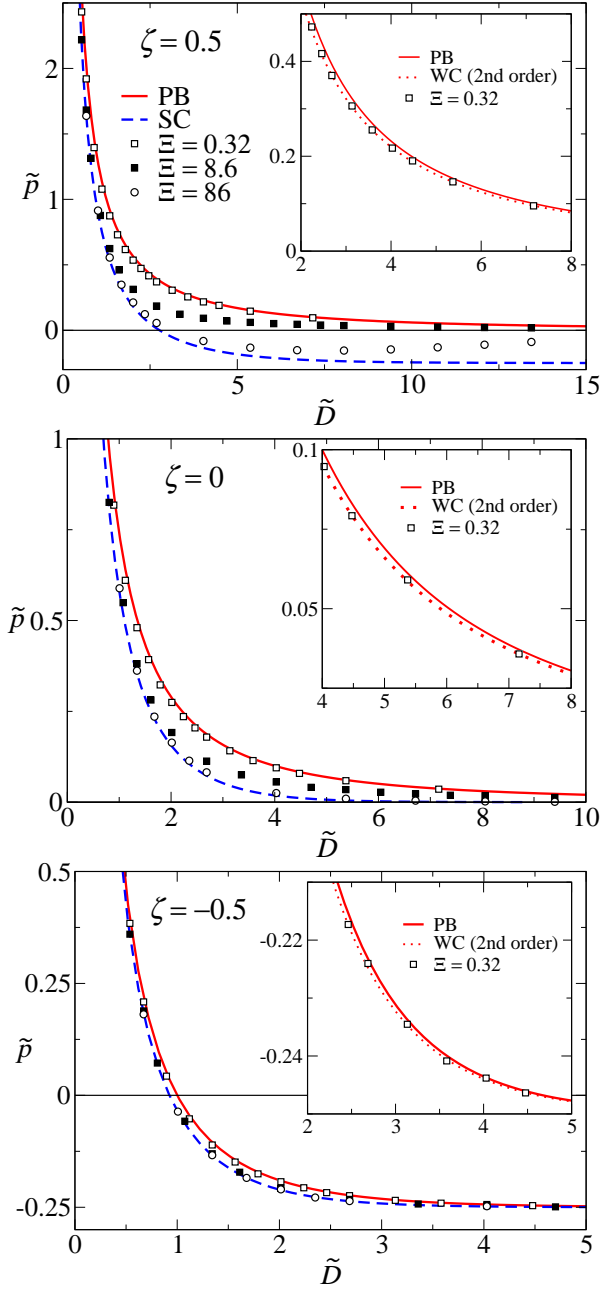

\centerline{\hspace{1.4ex}\psfig{figure=p2.eps,width=7.7cm}}
\centerline{\psfig{figure=p1.eps,width=7.9cm}}
\centerline{\psfig{figure=p3.eps,width=7.7cm}}
\caption{(Color online) Rescaled interaction pressure, $\tilde p$, as a function of the rescaled inter-surface distance, $\tilde D$, 
 for three different values of the asymmetry parameter $\zeta=0.5, 0$ and $-0.5$ (top to bottom).
 Solid lines represent the PB prediction, Eqs. (\ref{pres-pb-rep}) and (\ref{pres-pb-att}), dotted lines show the
second-order weak-coupling pressure (PB plus second-order corrections),  Eq. (\ref{secondord}), and  the dashed lines are the
SC prediction, Eq. (\ref{eq:P_SC}). Symbols correspond to MC simulations data at three different coupling parameters $\Xi=0.32$ (open squares), 
8.6 (filled squares) and 86 (open circles).
Insets show details at  small pressures along with the PB theory result as well as the second-order WC result for $\Xi=0.32$.}
\label{fig:sim_pressure}
\end{figure}

Next we consider the inter-surface pressure as obtained from 
the simulations (symbols in Fig. \ref{fig:sim_pressure}) as well as 
the PB theory, Eqs. (\ref{pres-pb-rep}) and (\ref{pres-pb-att}),  
and the SC theory, Eq. (\ref{eq:P_SC}) (solid and dashed lines, respectively).  
The PB result is expected to be valid for separations $\tilde D\gg \Xi$ \cite{Netz} (see also
Section \ref{sec:WC_crit}).
Therefore, for $\Xi = 0.32$ the PB line expectedly agrees nicely with the simulation data 
(open squares) in the whole range of separations shown in the figure. Upon closer
inspection, however, we find small deviations from the PB result as shown in the
insets in Fig. \ref{fig:sim_pressure} for all three values of the 
asymmetry parameter $\zeta$.  In this case, the fluctuation correction to the
mean-field pressure accurately compensates for these deviations and adding
the second-order correction to the PB pressure leads to a total
WC pressure, Eq. (\ref{secondord}) (shown as a dotted line in the inset) that 
matches the simulation data perfectly.  Note that the pressure changes can be drastic even on the WC level as $\zeta$ assumes different values. For $\zeta = 0.5$ and 0 the  WC pressure is strictly repulsive, while it turns attractive and leads to 
a bound state (zero pressure point) for $\zeta = - 0.5$.

In the intermediate regime of coupling parameters, the simulation results for the pressure 
are clearly bracketed by the two limiting analytical forms, given by the PB plus the second order correction and the SC expressions 
of the interaction pressure. The SC prediction is expected to be valid for separations $\tilde D\ll \sqrt{\Xi}$ as discussed in 
Section \ref{sec:SC_crit}. Consistently, the interaction pressure starts off close to the strong coupling limit at small separations 
and then smoothly converges to the weak coupling limit for larger separations. This is strictly true for $\zeta = 0.5$ and 0. In the case 
of $\zeta = -0.5$ the difference between the strong and weak coupling results
 for the rescaled interaction pressure is marginal  and the simulation data and the analytical results 
 nearly coincide for all rescaled separations $\tilde D=D/\mu$. 
We emphasize that the pressures and the density profiles are plotted here in {\em rescaled representation}; in actual units, 
Fig. \ref{fig:sim_pressure} corresponds to different ranges of separation, $D$, for the WC and SC
regimes as the Gouy-Chapman length, $\mu$, is typically very different 
between the two limits (small at high couplings and large at small couplings as may
be realized, {\em e.g.},  by changing the counterion valency at fixed surface charge densities and Bjerrum length). 

For large values of the coupling parameter the simulation results for the interaction pressure expectedly follow very closely the strong 
coupling prediction for a wider range of inter-surface separations.  
The correspondence between the SC theory and simulations is better for $\zeta = -0.5$ than for $\zeta = 0.5$, which can be again traced back to
the fact that the strong and the weak coupling results are very close to one another
for charged surfaces of opposite sign in the whole range of rescaled separations, whereas 
they differ significantly in the case of surfaces of equal sign. 
For intermediate and large couplings, we have not attempted to compare our data with the second-order WC
approximation as this approximation breaks down at the range of distances shown in the figures (Section \ref{sec:WC_crit}). 

Note also that in all cases considered here the theoretical and the simulated values of the interaction 
pressure converge for very small surface-surface separations. 
In fact, the SC and PB results coincide in the leading order as the rescaled distance, $\tilde D$, tends to zero, Eqs. (\ref{rep-limit1}) and (\ref{pSC_limit}), since both are dominated by the
osmotic pressure of counterions. The sub-leading corrections for very small $\tilde D$
are different in the PB and SC limits and on this level  the simulation data with finite $\Xi$ are generally expected to agree better with the SC prediction at small separations \cite{Netz}.

\begin{figure}[!h]
\centerline{\psfig{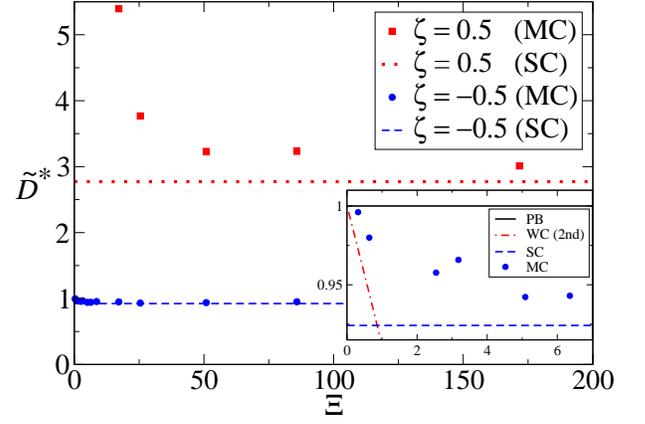}}
\caption{(Color online) Rescaled bound-state separation $\tilde D^\ast$ as a function of the coupling parameter $\Xi$ for different values of $\zeta$.
Main set: symbols are simulation data for $\zeta=0.5$ (filled squares) and $-0.5$ (filled circles). The SC results, Eq. (\ref{eq:D*_SC}),  for $\zeta=-0.5$ and 0.5 are represented by dashed and dotted lines, respectively. The inset represents the detailed view for $\zeta=-0.5$, where we also show the zeroth-order PB result, Eq. (\ref{eq:PB_a*}) (solid line),
and the result from the second-order WC approximation (dot-dashed line) 
obtained numerically from Eq. (\ref{secondord}). 
}
\label{fig:bound_Xi}
\end{figure}

\begin{figure}[!h]
\centerline{\psfig{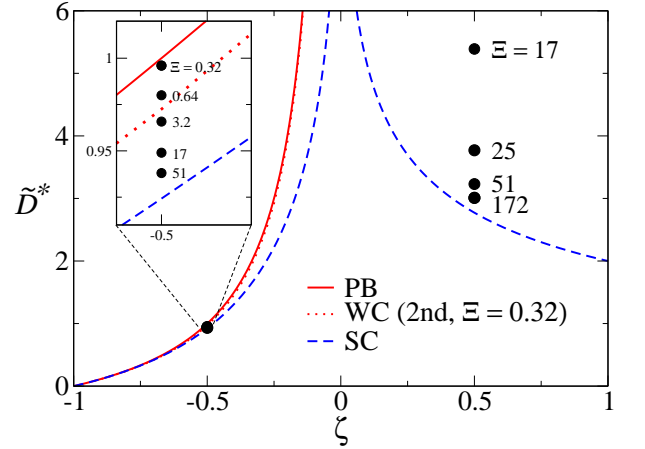}}
\caption{(Color online) Rescaled bound-state separation $D^\ast$ as a function of $\zeta$. Symbols are simulation data for different values of the
coupling parameter at $\zeta=0.5$ and $-0.5$.
The region around $\zeta = -0.5$ is expanded in the inset showing the crossover from the PB prediction (solid line) to the SC prediction
(dashed line) upon increasing the coupling parameter. }
\label{fig:bound_zeta}
\end{figure}

We now consider the simulated  bound-state separation, $D^\ast$, as a function
of the coupling parameter in Fig. \ref{fig:bound_Xi} (symbols). 
For $\zeta=0.5$, the PB theory ($\Xi\rightarrow 0$) gives only repulsion and predicts no bound state. 
While the SC theory predicts a closely packed bound state with $D^*$ given 
by Eq. (\ref{eq:D*_SC}), explicitly here $\tilde D^*\simeq 2.77$. 
As seen, by increasing the coupling parameter the simulation data for $D^\ast$ (filled squares) 
decrease monotonically and rather slowly converge to the SC prediction (dotted line). 
 The comparison is again worse for the $\zeta = 0.5$ than for the $\zeta = -0.5$ case 
 (dashed line and filled circles). 
 For  $\zeta=0.5$, even at $\Xi = 86$ the difference between the simulations and the analytical result is still close to
 10\%.  The deviations are quite pronounced for smaller values of the coupling parameter.
The opposite is true for $\zeta = -0.5$. Here, the simulation results are close to the strong-coupling analytical limit in the whole range of $\Xi$ 
values. For $\Xi < 4$ we can discern, see inset in Fig. \ref{fig:bound_Xi}, weak coupling behavior that starts off with the PB-predicted 
value $\tilde{D}^\ast = 1$ in the limit of $\Xi \rightarrow 0$ (Eq. (\ref{eq:PB_a*}), solid line), 
 that later follows the PB plus second-order corrections line (dot-dashed line) 
 and then rapidly approaches   the strong-coupling result $\tilde{D}^\ast \simeq 0.92$ (dashed line). 
As already noted the differences between strong and weak coupling in this case 
are marginal in the rescaled representation.

The dependence of $\tilde{D}^\ast$ on $\zeta$ in Fig.  \ref{fig:bound_zeta} complements the above observations.
The bound-state separation diverges for $\zeta = 0$ both in simulations as well as in the
  analytical limits. Here we reproduce the simulation results at $\zeta=0.5$ and $-0.5$, 
  which again clearly show the convergence to the SC result for $\zeta=0.5$ and
  the crossover from the PB result (solid line) to the SC result (dashed line)
  for  $\zeta=-0.5$  upon increasing the coupling parameter. 
The quantitative agreement between the simulations and the analytical results in the two limiting cases of PB and SC 
is excellent. Note here again that the second-order WC expansion  (dotted line in the inset) begins to fail at small separations 
as it deviates from the limiting PB results as well as from the simulation data. Since the PB pressure is zero 
on the PB phase boundary line, the validity of the second order WC correction can be assessed analytically only by performing
a two-loop calculation which goes beyond the scope of this paper.

\section{Conclusions}

To summarize, we have derived theoretical forms for the interaction pressure as well as the counterionic density profiles of asymmetrically charged planar surfaces with neutralizing counterions in between. Based on the field-theoretical methods we analyzed two different regimes of weak and strong coupling as defined by the electrostatic coupling parameter $\Xi$. The crossover between these two regimes
is studied {\em via} Monte-Carlo simulations. 

For small values of $\Xi$, the system is described very well by the {\em weak coupling} (WC) theory, that in the lowest  (zeroth) order coincides with the 
mean-field Poisson-Boltzmann (PB) result. The second order  
 of WC corresponds to a first-order loop expansion 
  and represents the contribution from correlated quadratic 
 fluctuations around the mean-field or saddle-point 
 solution. This second-order correction, which is proportional to $\Xi$, always lowers the interaction pressure between the surfaces and thus leads to an attractive contribution to the total interaction pressure. Since it corresponds to an expansion of the partition function around the mean-field saddle point it has to be smaller, in absolute terms, than the mean-field result. Net attraction given by 
 second-order fluctuations term (for interacting surfaces of equal sign) 
 is therefore inconsistent with the nature of the WC approximation.

For large values of the coupling parameter $\Xi$, the weak coupling approach breaks down and the virial expansion amounting to the  {\em strong coupling} (SC) approximation must be used. The SC approach is effectively a one-particle theory and takes properly into account the strong correlation and interaction of the counterions with external surface fields on the leading order \cite{Netz}. Following standard procedures, we derived an analytical expression for interaction pressure in the SC limit. The interaction pressure in this case is always lower than the PB result and can be negative (corresponding to a net attractive force) even for charged surfaces of equal sign.

We compared both our theories, {\em i.e.} WC with second-order corrections and SC, with Monte-Carlo simulations. We found very good agreement for both theories in their expected regime of validity. As expected, the WC approach is valid for separations $\tilde D\gg\Xi$. The second-order correction 
improves the small discrepancies between PB and MC results at large separations but it 
tends to fail for smaller distances when discrepancies get more pronounced. On the other hand, the SC theory describes the behavior perfectly at small separations $\tilde D\ll\sqrt{\Xi}$. For small enough coupling parameter $\Xi$, the validity of the PB approximation spreads to smaller separations, $\tilde D$, where PB and SC results nearly coincide (in the rescaled representation). Therefore, we may conclude that for sufficiently small $\Xi$, the PB result is valid on the whole interval $\tilde D$.

Note that the second-order WC correction term consistently diverges (toward 
large negative values) for small inter-surface separations, $\tilde D < 1$, irrespective of $\zeta$, which makes it in general inapplicable in this limit. The reason for this is simple. For small inter-surface separations the mean-field solution becomes more and more homogeneous, almost a constant, and the interaction free energy approaches its standard zero-frequency van der Waals form that diverges for small separations.

In the case of surface charges with equal sign, $\zeta>0$, the WC theory predicts no attraction and hence no bound state. The attraction and the corresponding bound state appear only for coupling parameters $\Xi$ that are large enough, as predicted by the SC theory. In the case of charged surfaces of opposite sign, $\zeta<0$, the attraction appears also in WC limit above a threshold value $\tilde D^\ast$ that represents the equilibrium surface-surface separation.

It is notable that  for charged surfaces of opposite sign, the WC analysis in general performs much better than for the surfaces of equal sign  and that the SC and the WC results are very close to one another for charged surfaces of opposite sign in the whole range of rescaled separations, whereas they differ significantly in the case of surfaces of equal sign. There is also only a marginal difference between the 
 PB and SC counterion density profiles in the rescaled representation for surfaces of opposite sign. A reasonable explanation for this would be in our opinion that for oppositely charged surfaces the counterions mostly feel the effect of the strong uniform external field provided by the surface charges, which acts similarly in the strong as well as the weak coupling limit. Thus the mean-field and the strong-coupling approaches should converge. 
 In the case of similarly charged surfaces, the mean-field theory depends more on the local counterion density whereas the strongly coupled counterions still feel mostly the external field. 
 Thus the difference between the WC and the SC frameworks in the $\zeta<0$ and $\zeta>0$ cases.

Our results support an emerging new paradigm, according to which the WC and the SC limit bracket the exact results for the interaction pressure between charged surfaces neutralized by  mobile counterions. They indeed provide quantitatively correct results for the interaction pressure in the limit of small and large inter-surface separations, while at intermeditae separations the exact results are always located between the two limits. It thus seems advisable that in analyzing the electrostatic interactions in colloidal systems one  always calculates both analytic limits, the WC as well as the SC, in order to get a good handle on the range of values that the interaction can assume for any value of the electrostatic coupling parameter. In future we intend to study the same system in the presence of added salt and dielectric discontinuities.

\section{Acknowledgements}

M.K. and R.P. would like to acknowledge the financial support by the Agency for Research and Development of Slovenia (Grants P1-0055(C), Z1-7171, L2-7080). This study was supported by the Intramural Research Program of the NIH, National Institute of Child Health and Human Development.
This research was supported in part by the National Science Foundation under Grant No. PHY05-51164.


\end{document}